\documentclass[reprint,amsmath,amssymb,aps,superscriptaddress,floatfix]{revtex4-1}

\usepackage{graphicx}
\usepackage{dcolumn}
\usepackage{bm}
\usepackage{hyperref}
\usepackage{siunitx}
\usepackage{color}
\usepackage{xspace}

\DeclareSIUnit\kT{$k_B T$}
\DeclareSIUnit\dyne{dyne}
\DeclareSIUnit\molar{M}
\newcommand{\kT}{{k_B T}}

\begin{document}

\title{Dynamic self-organized error-correction of grid cells by border cells }

\date{\today}

\author{Eli Pollock}
\email{epollock@mit.edu}
\affiliation{MIT, Brain and Cognitive Sciences, 43 Vassar St, Cambridge, MA 02142, USA}

\author{Niral Desai}
\email{nndesai01@gmail.com}
\affiliation{University of Texas at Austin, Department of Physics, 2515 Speedway, Austin, TX 78705, USA}

\author{Xue-xin Wei}
\email{weixxpku@gmail.com}
\email{Equal contribution}
\affiliation{Dept. of Statistics, Center for Theoretical Neuroscience, Columbia University, 
3227 Broadway, New York, NY 10027, USA}

\author{Vijay Balasubramanian}
\email{vijay@physics.upenn.edu}
\email{Equal contribution}
\affiliation{David Rittenhouse Laboratories, University of Pennsylvania, Philadelphia, PA 19004, USA}

\begin{abstract}
Grid cells in the entorhinal cortex are believed to establish their regular, spatially correlated firing patterns by path integration of the animal's motion.  Mechanisms for path integration, e.g. in attractor network models, predict stochastic drift of grid responses, which is not observed experimentally.  We demonstrate a biologically plausible mechanism 
of dynamic self-organization by which border cells, which fire at environmental boundaries, can correct such drift in grid cells.   In our model, experience-dependent Hebbian plasticity during exploration allows border cells to learn connectivity to grid cells.   Border cells in this learned network reset the phase of drifting grids.  This error-correction mechanism is robust to environmental shape and complexity, including enclosures with interior barriers, and makes distinctive predictions for environmental deformation experiments.   Our work demonstrates how diverse cell types in the entorhinal cortex could interact dynamically and adaptively to achieve robust path integration.
\end{abstract}

\maketitle

To survive, animals must have a way of knowing where they are within an environment.  A key component of this self-localization is  path integration \cite{Darwin1873}, namely the ability to continuously integrate  velocity  to keep track of position \cite{Wang2002, Etienne2004, Mcnaughton2006}.  What  neural mechanisms underlie  path integration?  Decades of research suggest that several brain regions may be involved, including hippocampus (which contains place cells \cite{o1971,OKeefe1976,o1978}) and entorhinal cortex (which contains grid cells \cite{Hafting2005,Fyhn2004} along with other spatially correlated neural types).  It is believed that the response of grid cells may be generated through path integration of an animal's trajectory through space.

In models of path integration, noise in neurons and inaccuracies in the integrator typically lead to accumulating errors in the location estimate \cite{Etienne2004, Mcnaughton2006}, 
sometimes causing substantial drifts over time in the grid response pattern \cite{Burak2009}.  However, such drifts are rarely seen experimentally grid cell firing. This suggests that the brain must contain mechanisms, perhaps involving environmental landmarks~\cite{Wang2002, Etienne2004, Mcnaughton2006,Mcnaughton1991,Touretzky1996, o2005,Skaggs1995,Giocomo2016environmental,Lever2002,Jacobs2003,Cheung2012,Savelli2017,Mulas2016}, to correct for drifts and to allow a stable representation of an animal's location.  

The boundaries of an environment are obvious landmarks, and it has been proposed that they provide cues for  path integration \cite{Mcnaughton1991, Touretzky1996, o2005, Hafting2005,Cheung2012,Krupic2016,Hardcastle2015}. Experimentally, it is known that environmental boundaries modulate the firing properties of place cells~\cite{O1996} and grid cells~\cite{Stensola2015shearing,Krupic2015}.   Recently, a new class of boundary-sensitive neurons was discovered in the entorhinal cortex and nearby regions \cite{Solstad2008, Lever2009, Savelli2008,Stewart2014,Muessig2015}, and there is evidence that these ``border cells'' may be involved in error correction of grid cell responses \cite{Hardcastle2015,Giocomo2016environmental}.   

The authors of ~\cite{Hardcastle2015} proposed a model with a simple hand-crafted pattern of connectivity between border cells and grid cells to correct drifts of grid firing patterns in a square environment.  This pioneering effort, however, did not address how border-to-grid connectivity might be learned to provide error correction in environments of different, and more complex, shapes.  This is an important challenge because environments change, animals move to new locations, and natural habitats often have complex geometries with internal barriers and passageways.

Here, we develop a modeling framework in which connections between border cells and grid cells are self-organized to provide an error correction signal for grid cells while the animal freely explores an environment.  In our model, grid responses are generated by attractor dynamics, similar to previous models \cite{Fuhs2006, Burak2009, Bonnevie2013,Couey2013}, but are modified by a learned interaction with border cells. We show that the resulting border-grid network dramatically reduces drift in  the grid system. Our proposed mechanism can correct grid drifts across environments with different shapes and complexities, for example when internal barriers are introduced, enabling a stable representation of self-location in ethologically relevant situations.   Our model  also makes distinctive predictions for grid drift after changes in the spatial environment.  Together, these results provide insights into the use of environmental boundaries, and landmarks in general, to achieve robust path integration.  This paper expands work  presented at the Computational and Systems Neuroscience meeting (CoSyNe) in February 2017.

\section*{Results}
\vspace{-0.15in}

\subsection*{Grids drift linearly with time in an attractor model} 
\vspace{-0.15in}
We constructed a continuous attractor model of the grid system following \cite{Burak2009,Couey2013} (Methods).  In this model, grid cells are subject to external excitation modulated by movement velocity and inhibit each other recurrently.   The network self-organizes into a grid pattern of activity on the neural sheet. The pattern is translated on the neural sheet as the animal moves, with the consequence that a given grid cell fires when the animal is physically located in a triangular grid of locations in the environment  (Fig. 1a, blue points or red points).   
We found that the firing pattern of individual grid cells drifts in space over time (Fig. 1a, blue to red gradient).   The average cross-correlation of grid firing patterns at different times was used to quantify the mean squared grid drift (Methods) as the simulated animal explored the environment (trajectories in Methods).   We found that the mean squared drift increased linearly with time, as expected for a two dimensional random walk (Fig. 1b).

\begin{figure}
\centering
\includegraphics[
keepaspectratio,width=\linewidth]{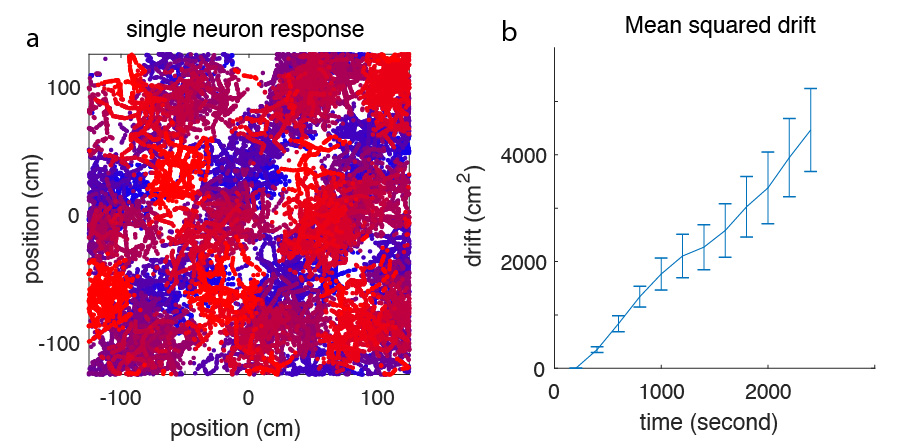}
\caption{Grid drifts in a continuous attractor network model. a) Response of a single model neuron in a continuous attractor network model for the grid system.  At a given time a grid cell responds if the animal is located near the vertices of a triangular lattice in space (blue points, burgundy points, or red points).  However, as time passes the grid pattern shifts in space (blue = early times, burgundy = intermediate times, red = late times).   Averaging over time can destroy the grid-like appearance of the firing pattern.   b) Drift, quantified as mean squared shift between triangular grid patterns, increases linearly in time (see Methods).  Error bars are standard error over 50 replicate simulations of animal trajectories.
}
\end{figure}

\subsection*{Modeling responses of border cells}
\vspace{-0.15in}
We considered that the drift observed in Fig.~1 might be corrected if the grid attractor network interacted appropriately with landmarks, notably boundaries.  To this end, we constructed a model population of border cells \cite{Solstad2008, Lever2009, Savelli2008,Stewart2014,Muessig2015} which are known to fire near environmental boundaries (Methods).   Consistently with experiment  \cite{Solstad2008} we assumed that border cells fire over scales that range between half to all of the sidelength of rectangular boundaries  (125 cm to 250 cm in our simulation; Fig.~2).   While we assumed that border cells have some bias towards lying on only one wall in the environment, we also allowed them to wrap around corners (Fig.~2), consistent with the observation that border cells frequently have a ``dominant wall" but are not necessarily wall-specific \cite{Solstad2008}. We also assumed that border cells map to new environmental configurations by firing in the proximity of boundaries in a fixed allocentric direction \cite{Lever2009, Solstad2008}.    Thus, if the environment was deformed or a new internal barrier was introduced, the model border cells continued to fire along segments of the new boundary with same absolute orientation and over similar length scales (Methods).

\subsection*{Self-organization of border-to-grid connectivity}
\vspace{-0.15in}
In our model,  border cells develop connectivity with grid cells via experience-dependent Hebbian plasticity~\cite{Hebb1949}.  When an individual border cell fires, its synaptic weight with a synchronously firing grid cell increases.   Specifically,  the connection strength between the $i^{th}$ pre-synaptic neuron (border cell) and the $j^{th}$ post-synaptic neuron (grid cell),  $W_{ij}$, is updated 
according to a learning rule~\cite{Hebb1949},
$
W_{ij}^{t} = W_{ij}^{t-1}  + \gamma x_i y_j \, ,
$
where $\gamma$ is a learning rate and  $x_i, y_j = 1,0$ denote firing or silence of border cell $i$ and grid cell $j$ respectively (Methods).
Divisive normalization is then applied to maintain the overall connectivity strength between border and grid populations~\cite{Miller1994}, 
i.e.,
$
W_{i,j}^{t} \to W_{i,j}^{t} / \sum_j W_{i,j}^{t} .
$
Such normalization is needed for the stability of Hebbian learning (e.g.,~\cite{Miller1994}).

The border-to-grid connectivity pattern stabilizes as the animal gains experience in an environment, and strong connections form between each border cell and grid cells that fire at locations within its spatial firing field (Fig.~2 insets).  Grid cells whose spatial phases are shifted by a grid lattice vector will have identical connection strengths, leading to the connectivity patterns in Fig.~2. Our model predicts that  the spatial phase of a grid cell will be correlated with its connection strength with a  border cell in a manner that depends on the spatial boundary geometry and on the border field location.   These predictions can  be tested experimentally by reconstructing the functional connectivity of grid cells and border cells via simultaneous recording of populations using multi-electrode recording or calcium imaging.

\begin{figure}
\centering
\includegraphics[
keepaspectratio,width=.6\linewidth]{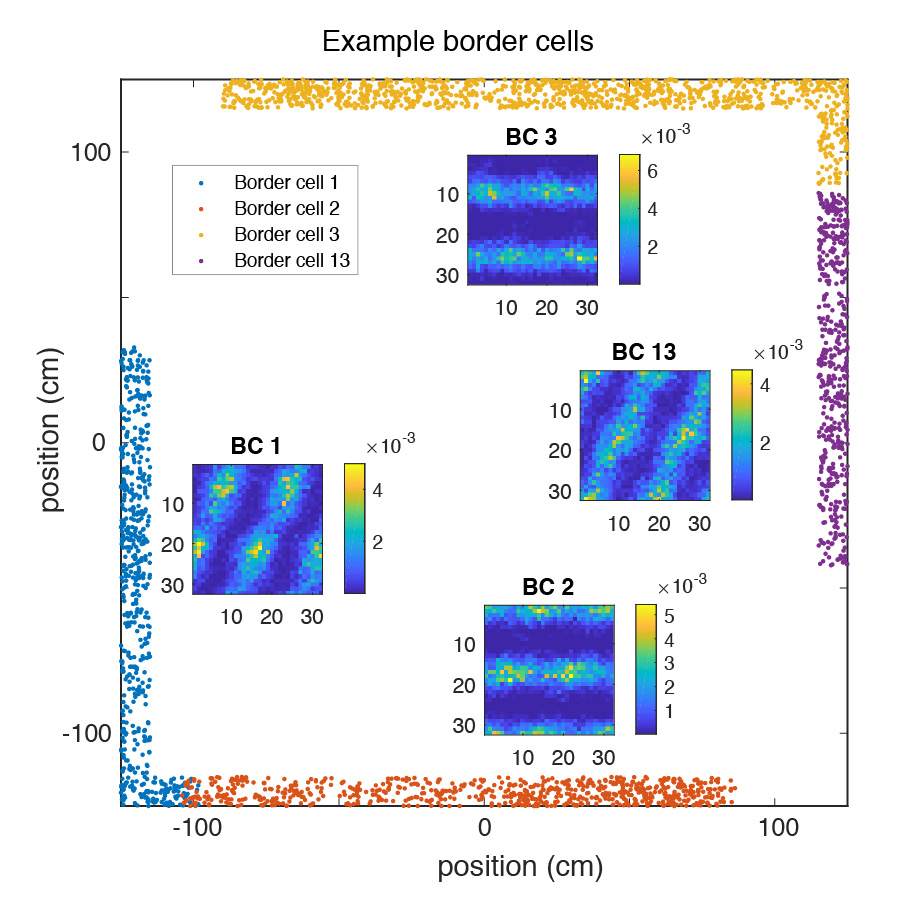}
\caption{Border cell firing and connectivity. Four example border cells and their firing fields are shown around the boundary of the spatial enclosure (large square).   Insets:  Heatmap indicates the strength of connections  between indicated border cell and grid cells organized by spatial phase on the neural sheet.   The periodic connectivity pattern occurs because grid cells whose phase differs by a grid lattice vector respond in the same physical locations.}
\end{figure}

\subsection*{Error-correction in simple environments }
\vspace{-0.15in}
Intuitively, the mechanism described above will store the relative phase of grid and border cells in the strength of synapses connecting them.  Thus,  in a mature network, border cell firing should on average reinstate the phase of grid cells that may have drifted.   To test this we allowed grids to form with and without simultaneous development of border-grid connectivity in a square environment. For the model with connectivity, the corrective input to each grid cell was a sum over the activity of border cells weighted by the border-grid synaptic strength (Methods).   We quantified the displacement of grid firing fields relative to their initial location in terms of a mean squared grid drift (Methods).  Learned border-grid connectivity dramatically attenuated drift in the square environment (Fig.~3a,b).   Grid stabilization was apparent in single cell firing patterns (compare blue and red points in Fig. 3a vs. Fig. 1a), and also persisted in circular environments (Fig.~3c,d).  Thus, grid drift can be corrected by interaction with border cells in simple convex environments where each border field is attached to a single, compact region of the boundary.

\subsection*{Error correction in complex environments}

Ethologically relevant environments can have complex, non-convex  shapes, sometimes with internal barriers.  In such spaces border cells may fire in multiple disjoint regions.  For example, in the presence of an internal barrier (Fig.~3e), border cells can respond both to the barrier and to walls parallel to the barrier \cite{Solstad2008, Lever2009}.    In such situations grid cells with different phases can respond at the same time as a given border cell with disjoint response fields.
 This possibility complicates the grid phase re-setting  by border cell input that is required to correct drift.   We therefore tested our mechanism in an enclosure with an internal barrier (Fig.~3e), with border cells assigned to respond in the same allocentric direction at the barrier and boundary walls (Methods), consistently with experiment \cite{Solstad2008, Lever2009}.  Hebbian plasticity (Methods) caused border cells to form strong connections with  concurrently firing grid cells, so that cells with multiple border fields synapsed with grid cells with different phases.  Again, grid drift was strongly attenuated by interaction with border cells (Fig.~3f).   Error-correction occurs robustly so long as accumulated drift is small, because in this case the correct grid cells will be primed to fire near the internal barrier and near the boundary, albeit a little out of phase.  Thus,  border cell firing will simply reinstate the proper phase in the grid cell, causing it to ``snap back'' to the right spatial pattern.    If the accumulated drift were large, error correction could fail, but this is not the case when there is regular contact with boundaries.

\vspace{-0.15in}
\begin{figure}
\centering
\includegraphics[keepaspectratio,width=\linewidth]{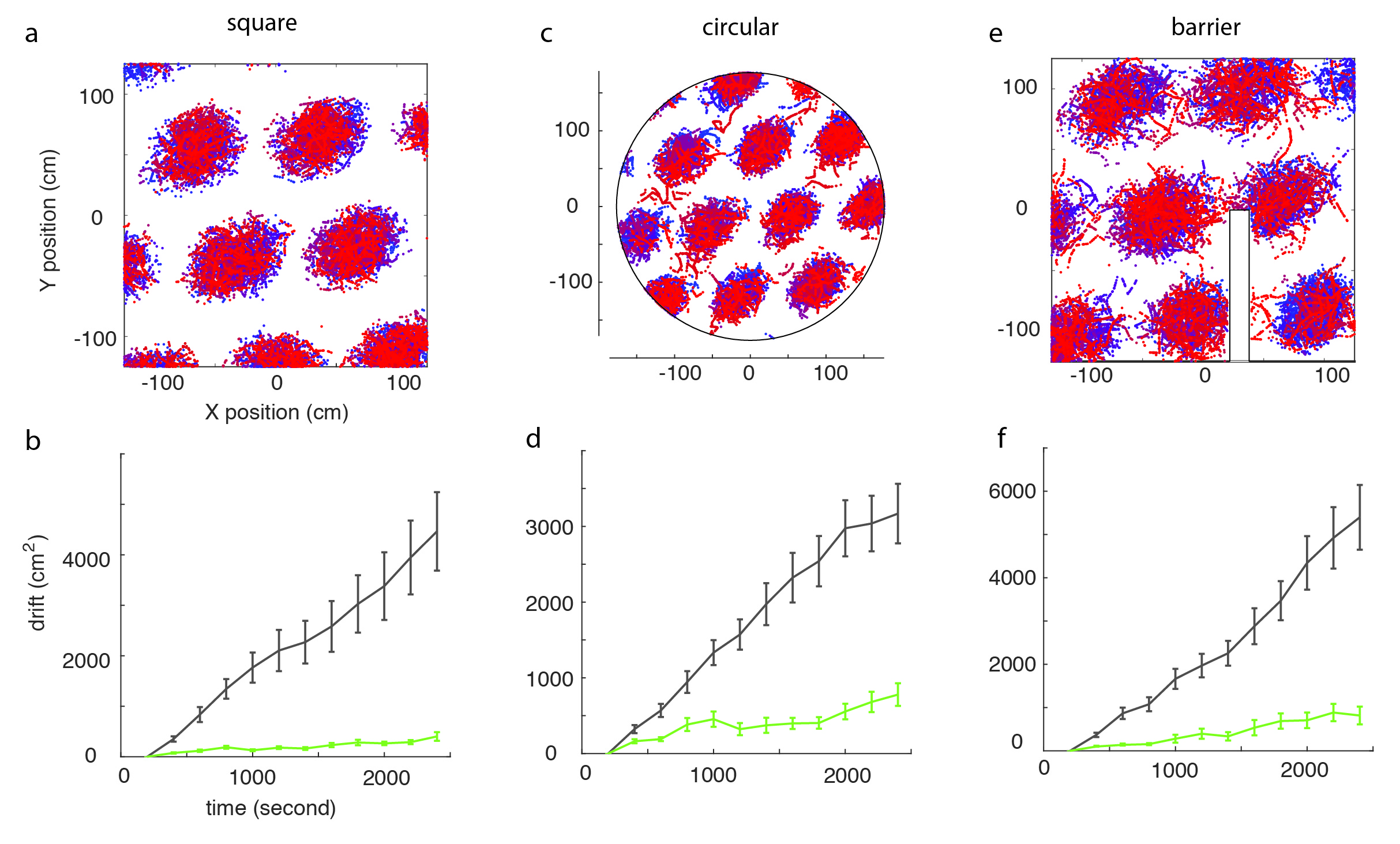}
\caption{Correction of grid drift by learned border input. (a-b) Square environment.  Top:  firing pattern for a single grid cell with error correction by border inputs,  blue dots = early times, red dots = late times. Bottom: Grid drift as a function of time with (green) and without (black) border cell inputs.  
Error bars = standard error over 50 replicate simulations.  
(c-d) Similarly for a circular enclosure. (e-f) Similarly for a square with an internal barrier.}
\end{figure}

\subsection*{Error correction in changing environments }
\vspace{-0.15in}
\begin{figure*}
\centering
\includegraphics[keepaspectratio,width=0.99\linewidth]{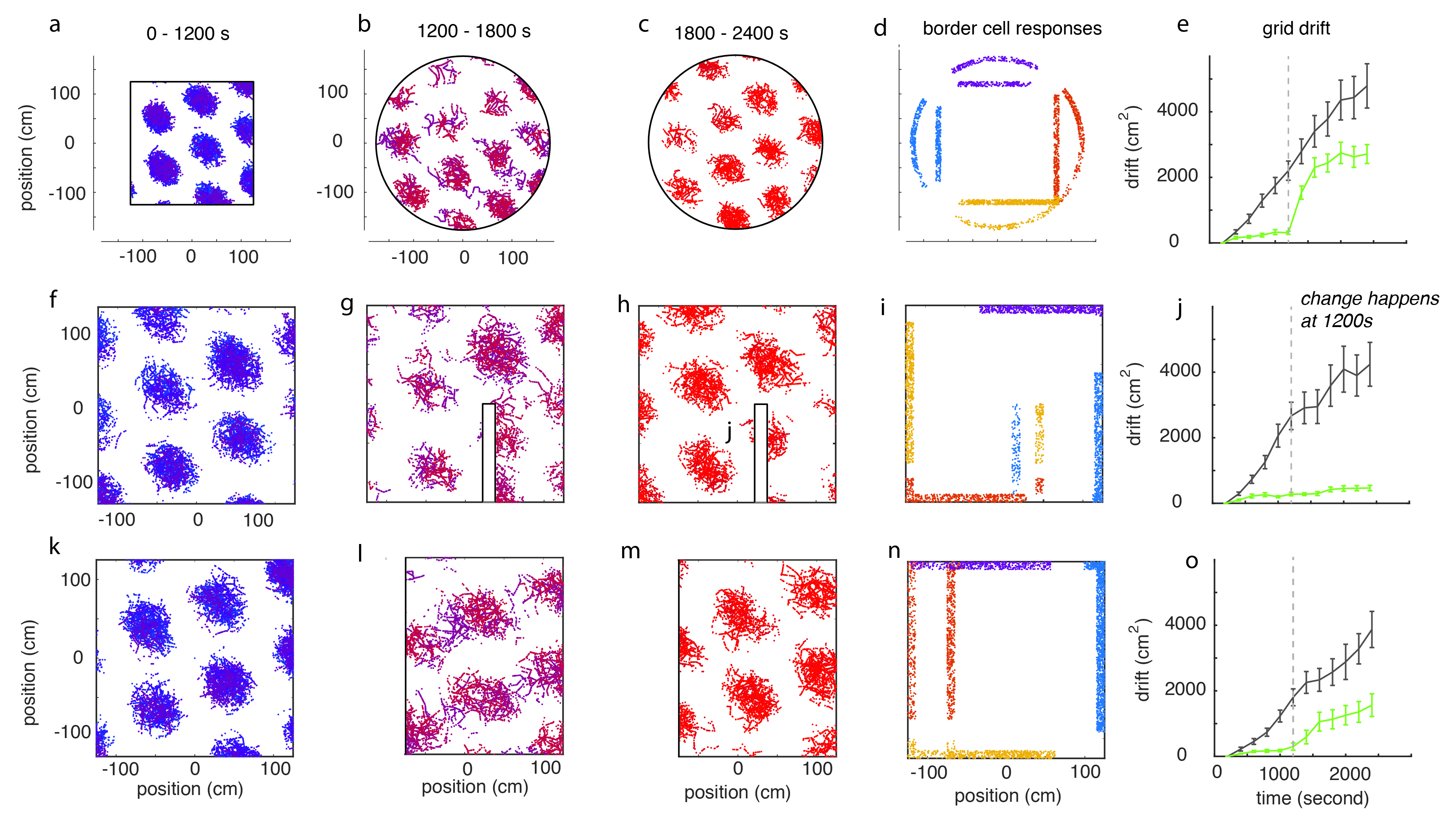}
\caption{Correction of grid drift in changing environments. (Top Row, a-e) Square-Disc deformation.   The square environment  in (a) is deformed to a  disc in (b) at 1200s leading to disruption of the grid pattern which then stabilizes in (c) at late times.  Border firing fields in (d) map allocentrically between  enclosures.   Without border inputs grids drift similarly before and after the deformation.  Self-organized error correction from border cells leads to much lower drift before the deformation, a transient increase in drift after the deformation, and re-stabilization as the changed environment is re-learned (error bars = standard error over 50 replicates).  (Middle Row, f-j) Insertion of an internal barrier.    (Bottomg Row, k-o) Square-Rectangle deformation.
}
\end{figure*}

Real-world navigation occurs in a dynamic context where environments change, and where objects, barriers, and other animals enter or leave a space.   To test whether border cell input  can stabilize grids during and after a spatial deformation we first  morphed a square enclosure into a disc, while mapping border cells so that they responded to boundaries in the same allocentric orientations in both shapes (Fig.~4, top row).    After the deformation,  stability of the grid pattern was temporarily disrupted and the rate of drift increased dramatically.  In fact the  drift  was higher than it would have been without an error-correction mechanism.  However, as the animal continued to explore, ongoing plasticity reorganized border-grid connections to be appropriate for the new enclosure, once again stabilizing the grid.  We also tested two milder deformations -- the sudden insertion of a narrow internal barrier (Fig. 4, middle row), and compression of the square to a rectangle (Fig.~4, bottom row).   Again the grids were transiently disrupted and then stabilized, but the effects were weaker than after the more dramatic square-to-disc deformation.   These results highlight the importance of ongoing learning and self-organization in correcting errors and maintaining stability  in path integration mechanisms operating in dynamically changing environments.

\section*{Discussion}
\vspace{-0.15in}

We have demonstrated  that border cells that self-organize connectivity with grid cells can correct accumulating positional drift, even in complex geometries that change in time.  We expect that such interactions  with boundaries will also be able to correct path-integration errors in other models of grid formation \cite{hasselmo2007grid} and boundary-related responses \cite{O1996,Barry2006boundary,Lever2002,Lever2009}.   Indeed, such anchoring may be generally necessary for reliable evidence accumulation with noisy neurons~\cite{Faisal2008}.

Many authors have noted that diverse sources of sensory and non-sensory information including landmarks and boundaries must be merged to maintain reliable spatial representations during navigation \cite{Skaggs1995,Gothard1996,Jacobs2003,Cheung2012,Perez2016,Raudies2016,Raudies2015}.    In  the grid system such external information must be passed in without interrupting the internal dynamics, so that  grid cells can continue to perform path integration while being informed by external cues.  Our results show that boundary information, at least, will be effectively injected if border cells and grid cells that fire in phase develop strong synapses.  Perhaps this simple mechanism provides a prototype for more general cue integration in the spatial navigation systems of the brain.

Several authors have explored the learned use of landmark and sensory information to correct errors, e.g., in the head-direction system \cite{Skaggs1995,Zhang1996,Knierim1996}, and in spatial representations maintained by robots \cite{Mulas2016}.   One  study  showed that learned connections  between ``landmark cells" and grid cells would lead to  grid firing deformations consistent with experiment \cite{Ocko2018}.  Meanwhile, \cite{Santos2017} explored how learned modifications of border-grid synapses affect the error-correction scenario of \cite{Hardcastle2015}.   We add to this literature by showing how a particular kind of landmark -- a boundary -- can provide cues for error correction even in complex and changing environments.

Our model makes several testable predictions.  First,  border-grid connection strengths will be correlated with grid cell spatial phase in a boundary geometry and  border field location dependent manner.   Second, the connectivity pattern will adapt to recent experience.  Third, grid firing patterns will become animal trajectory dependent when an environment changes shape because grids remain tethered to the now-deformed boundary.   Very recently, the authors of \cite{Keinath2017}  presented direct evidence for such trajectory dependence, and argued that averaging over trajectories partially reproduces the apparent effect of grid rescaling after environmental deformation \cite{Barry2007,Stensola2012}.

It is challenging to stabilize path integrators in complex environments. Our paper proposes a simple solution that might be realized in the brain, and will be useful in the design of robots learning to navigate rich terrains.

\begin{acknowledgments}
\vspace{-0.15in}
This work was partially supported by  the Honda Research Institute and NSF grant PHY-1734030 (VB), NSF grant PHY-1607611 (Aspen Center for Physics), the Penn Vagelos Molecular Life Sciences Program (EP, ND) and the NSF NeuroNex Award DBI-1707398 (X-XW).
\end{acknowledgments}

\section*{Methods}
\vspace{-0.15in}

\subsection*{Attractor network model for the grid cells}
\vspace{-0.15in}
We implemented the grid cell attractor model in \cite{Burak2009}, modified as in \cite{Couey2013} to simplify neural connectivity.  A two-dimensional sheet of $N \times N$ neurons with periodic boundary conditions follows an update equation
\[
\tau \frac{ds_i}{dt} + s_i = g\Big[\sum_j {M_{ij} \, s_j} + I + \alpha \, v_t \,  cos(\theta_t-\theta_i) + C_{i}(t) \Big]_+ 
\]
Here, $s_i$ is the activity level of neuron $i$, $g$ is a gain parameter, $\tau$ is the time constant, $I$ is a constant external input, $\alpha$ defines the velocity gain, $v_t$ and $\theta_t$ are the speed and head direction at time $t$, and $\theta_i$ is the preferred direction of neuron $i$. Preferred direction is chosen from the four cardinal directions by tiling the neural sheet with $2 \times 2$ squares containing up, down, left, and right preferences in a fixed pattern. The term $C_{i}(t)$ is the corrective input from the border cells to grid cell $i$ at time $t$, and is defined later. In simulations without  border cells, $C_i(t) = 0$. The notation $\big[.\big]_+$ indicates a rectified linear function.

Connection strengths between neurons are determined by their distance on the neural sheet.  Neurons within a disc of radius $R$ have a constant inhibitory weight $M_0$, while neurons outside have zero weight.  The disc is slightly offset by a distance $l$ in the neuron's preferred direction.  All told the connection weights between neurons at $(x_i,y_i)$ and $(x_j,y_j)$ are given by $M_{ij} = M_0  \times  \Theta\Big(R - \sqrt[]{(x_i-x_j-l \cos{\theta_i})^2 + (y_i - y_j - l \sin{\theta_i})^2}\Big) $, where $\Theta$ is the Heaviside step function. For parameter values, we used $N=32$, $\tau=10\, {\rm ms}$, ${\rm dt}=1\,{\rm ms}$, $g=1$, $I=3$, $\alpha=2$, $M_0=-0.05$, $R=13$, and $l$=2.

\subsection*{Navigation Simulation}
\vspace{-0.15in}
Animal speed, head direction, and border interactions  were generated by simulating an animal's path. 

\vspace{0.1in}
\noindent {\bf Environments: }  The basic enclosure was a square with side length 250 cm.  Circular environments were chosen to exactly circumscribe the square.
The barrier environment had an internal wall of width 20 cm and length 125 cm,  touching the southern wall at a location 30 cm to the right of the midline of the square enclosure.

\vspace{0.1 in}
\noindent{\bf Paths: }
Trials ran for 2400 seconds of simulation time. Animals moved at a constant speed of 1 m/s, starting from the center of the environment in a random direction. Every  0.1 seconds of simulation time (100 time steps), we chose a new direction by drawing from a Gaussian distribution with a standard deviation of one radian.   Whenever the animal reached a boundary we randomly reassigned the head direction for the next step so that the trajectory remained within the enclosure.

\subsection*{Border cell assignment}
\vspace{-0.15in}
 In the square enclosure  each border cell was centered randomly along the middle 50\%  of a randomly chosen wall.  The border field length was chosen uniformly from the interval $(L,L/2)$, where $L$ is the wall  length.   The average border cell thus took up 3/4 of a wall and was defined to have a field width of 10 cm, consistently with observation \cite{Solstad2008}. Border cells whose ends went beyond the wall on which they were centered wrapped around corners to other walls.    Thus border cells have a bias towards lying on a single ``dominant wall'', but are not wall-specific.   We used 16 border cells in all simulations.

To create border cells for circular environments, we projected border fields from the square environment out to the circumscribing circle. For rectangular environments, we rescaled border fields to occupy the same proportion of space on walls that were compressed. Finally, when introducing a barrier, we defined border cells so that they responded in the same allocentric direction at the barrier and the boundary walls.   For example, if a border cell fired when the animal was at the lower part of the right wall, it also fired at the corresponding location on the left face of a barrier inserted into the environment.

\subsection*{Learning border-to-grid connectivity}
\vspace{-0.15in}
The update rule and divisive normalization for border-grid connection weights $W_{ij}$ are given in the main text.   Grid and border cells spiked with Poisson statistics.  In each 1 ms timebin, border cell firing probability was 0.01 when the animal was inside the firing field and 0 otherwise.   For grid cells, we followed a hybrid of  \cite{Burak2009} and \cite{Couey2013} and defined firing probability in 1 ms timebins as the synaptic  activation (right hand side of the attractor equation) times 0.118.
The corrective input to the $j^{th}$ grid cell  was a weighted sum of border cell activities
$$
C_j = \beta \sum_{i} W_{ij} \, x_i
$$
with $\beta=200$ and $x_i = 1,0$ if a border cell fires or not.

\subsection*{Quantifying grid cell drift}
\vspace{-0.15in}
To quantify drift of grid response fields, we recorded locations of the Poisson spikes of a single grid cell.  Simulations were broken into 12 windows of 200 seconds each that were long enough to explore the environment, but short enough that the grids remained stable. In each window, we created a two-dimensional histogram of spike locations, with bin widths of 1 cm. We then computed a cross-correlation matrix of time-adjacent histograms. The drift between time windows was calculated as the displacement from the origin of the most central peak of the cross-correlogram. By adding drift vectors across subsequent time windows, we quantified drift in single trials.  To generate drift plots (Fig. 1b, Fig. 3 bottom, and  Fig. 4 right, we averaged the squared magnitude of drift across 50 trials in each time window.


%


\end{document}